\documentclass[letterpaper, 10 pt, conference]{IEEEtran}
\IEEEoverridecommandlockouts
\usepackage{cite}
\usepackage{amsmath,amssymb,amsfonts,amsthm}
\usepackage{algorithmic}
\usepackage{svg}
\usepackage{graphicx}
\usepackage{textcomp}
\usepackage{xcolor}
\usepackage{soul}
\usepackage{hyperref}
\usepackage{cleveref} 
\usepackage{subfigure}
\usepackage{float}

\newcommand\rColor{black} % Color of revision text
\newcommand{\revised}[1]{\textcolor{\rColor}{#1}}

\def\BibTeX{{\rm B\kern-.05em{\sc i\kern-.025em b}\kern-.08em
    T\kern-.1667em\lower.7ex\hbox{E}\kern-.125emX}}

\newtheorem{theorem}{Theorem}
\newtheorem{lemma}{Lemma}
\newtheorem{remark}{Remark}

\newtheorem{assumption}{Assumption}

\usepackage[left=0.75in, right=0.75in, bottom=0.75in, top=1in]{geometry}

\newcommand{\dispdot}[2][-.3ex]{\dot{\raisebox{0pt}[\dimexpr\height+#1][\depth]{$#2$}}}% \dispdot[<displace>]{<stuff>}
\newcommand{\sspace}{\mkern9mu}
\newcommand{\nspace}[1]{\mkern#1mu}

\newcommand{\nquad}[1]{\hspace*{#1em}\ignorespaces}

\newcommand{\norm}[1]{\left\Vert #1 \right\Vert}

\newcommand{\Cs}{\mathcal{C}_{\rm S}} % safe set 
\newcommand{\Cb}{\mathcal{C}_{\rm B}} % backup set 
\newcommand{\Cbi}{\mathcal{C}_{\rm BI}} % backward image set 
\newcommand{\Cd}{\mathcal{C}_{\rm D}} % backward image set 
\newcommand{\Ci}{\mathcal{C}_{\rm I}} % backward image set 
\newcommand{\ub}{\boldsymbol{u}_{\rm b}} % backward control law

\newcommand{\phidg}[2]{\boldsymbol{\phi}^{d} (#1, #2)}
\newcommand{\phidb}[2]{\boldsymbol{\phi}^{d}_{\rm b} (#1, #2)}

\newcommand{\phinb}[2]{\boldsymbol{\phi}^{n}_{\rm b} (#1, #2)}

\newcommand{\phinom}{\phinb{\tau}{\boldsymbol{x}}}
\newcommand{\phinomT}{\phinb{T}{\boldsymbol{x}}}

\newcommand{\stmnom}{\boldsymbol{\Phi}_{\rm b}^{n}(\tau,\boldsymbol{x})}
\newcommand{\stmnomT}{\boldsymbol{\Phi}_{\rm b}^{n}(T,\boldsymbol{x})}
\newcommand{\jac}{F_{\rm cl}}

\newcommand{\phid}{\phidb{\tau}{\boldsymbol{x}}}

\newcommand{\phidT}{\phidb{T}{\boldsymbol{x}}}

\newcommand{\tbSet}{\tau \in [0,T]}
\newcommand{\tb}{\tau}

\newcommand{\xzero}{\boldsymbol{x}_0}

\newcommand{\w}{\boldsymbol{\omega}}
\newcommand{\J}{\boldsymbol{J}}

\begin{document}

\title{\LARGE \vspace{-.65cm} \textbf{Disturbance-Robust Backup Control Barrier Functions: \\ Safety Under Uncertain Dynamics}}

\author{David E.J. van Wijk$^{1}$\thanks{Approved for public release. Distribution is unlimited. Case \#AFRL-2024-4823. This work was sponsored by AFRL under the STARS Seedlings for Disruptive Capabilities Program including contracts with UDRI (\#FA8650-22-C-1017). S. Coogan was supported in part by the NSF, award \#2219755.
% The views expressed are those of the authors and do not necessarily reflect the official guidance or position of the United States Government, the Department of Defense or of the United States Air Force. 
\newline \text{   } $^{1}$Aerospace Engineering, Texas A\&M University, College Station TX 77845, U.S.A, \texttt{\{davidvanwijk,mmajji\}@tamu.edu}.}, Samuel Coogan$^{2}$\thanks{$^{2}$Electrical and Computer Engineering, Georgia Institute of Technology, Atlanta GA 30332, U.S.A, \texttt{sam.coogan@gatech.edu}.}, Tamas G. Molnar$^{3}$\thanks{$^{3}$Mechanical Engineering, Wichita State University, Wichita KS 67260, U.S.A, \texttt{tamas.molnar@wichita.edu}.}, Manoranjan Majji$^{1}$, and Kerianne L. Hobbs$^{4}$\thanks{$^{4}$Safe Autonomy Lead, Air Force Research Laboratory, Wright-Patterson Air Force Base OH 45433, U.S.A, \texttt{kerianne.hobbs@afrl.af.mil.}}}

\maketitle

\begin{abstract}

Obtaining a controlled invariant set is crucial for safety-critical control with control barrier functions (CBFs) but is non-trivial for complex nonlinear systems and constraints. Backup control barrier functions allow such sets to be constructed online in a computationally tractable manner by examining the evolution (or flow) of the system under a known backup control law. However, for systems with unmodeled disturbances, this flow cannot be directly computed, making the current methods inadequate for assuring safety in these scenarios.
To address this gap, we leverage bounds on the nominal and disturbed flow to compute a forward invariant set online by ensuring safety of an expanding norm ball tube centered around the nominal system evolution. We prove that this set results in robust control constraints which guarantee safety of the disturbed system via our \textit{Disturbance-Robust Backup Control Barrier Function (DR-bCBF)} solution. 
The efficacy of the proposed framework is demonstrated in simulation, applied to a double integrator problem and a rigid body spacecraft rotation problem with rate constraints.

\end{abstract}

\section{Introduction}

\textit{Control barrier functions (CBFs)} \cite{ames_2017}, are a popular approach to assuring safety of autonomous systems by encoding safety into existing controllers and providing sufficient conditions for forward invariance of safe sets. However, obtaining safe sets for which every state has a safe control action (a.k.a. \textit{controlled invariant} sets) is difficult for high-dimensional systems\revised{, especially when considering input bounds}. Additionally, dynamics models are seldom perfect. In this letter we seek to solve both of these problems simultaneously. 

To address the problem of controlled invariance, we adapt the {\em backup set method} \cite{gurriet_online_2018, gurriet_scalable_2020, gurriet_thesis} based on online backward reachability. This method establishes a controlled invariant safe set \textit{implicitly} using the flow of the system under a prescribed backup control law. This approach is computationally tractable even for complex systems. For affine nonlinear systems, this technique generates linear control constraints which can be used to efficiently solve for point-wise optimal control signals for an arbitrary primary controller. 

The second problem we address is that of model uncertainty, \revised{
which has been studied extensively in the CBF literature.
Robust methods \cite{jankovic_robust_2018,garg_robust_2021,breeden_robust_2023,shaw_cortez_control_2021} typically rely on accounting for worst-case disturbances through an upper disturbance bound, and these methods can be made less conservative via disturbance estimation \cite{das2023robustcontrolbarrierfunctions}. The notion of input-to-state safety, defined first in \cite{issf_og} and extended for CBFs in \cite{Issf_ames}, provides a technique for handling input disturbances and has been successfully applied in multiple scenarios \cite{issf_application_ames,TEZUKA_issf}. Adaptive CBF methods have been shown to assure safety in the presence of parametric dynamics uncertainty \cite{ames_adaptiveCBF,xiao_adaptiveCBF} and a robust adaptive CBF extension can reduce conservatism and closed-loop chattering \cite{lopez_adaptive_robustCBF}. Learning \cite{Taylor2019LearningFS,lindemann2024learningrobustoutputcontrol} and data-driven \cite{taylor_robust_dataDriven,fernando_data-driven,emam_data_driven} approaches have also been developed to account for uncertainty in dynamics, state, or both. Lastly, for mixed-monotone systems their decomposable structure can be exploited to produce robustly forward invariant sets \cite{abate_robust_monotonicity}. 
While these approaches present viable solutions to addressing model uncertainty, they assume that a controlled invariant safe set can be found explicitly---a strong assumption for many systems and safety constraints. Works \cite{abate_enforcing_2020} and \cite{cosner_measurement-robust_2021} do not make this assumption, but the former is specific to mixed-monotone systems, and the latter assumes perfect dynamics knowledge and bounded measurement error.}
\begin{figure}[t]
    \centerline{\includegraphics[width=.86\columnwidth]{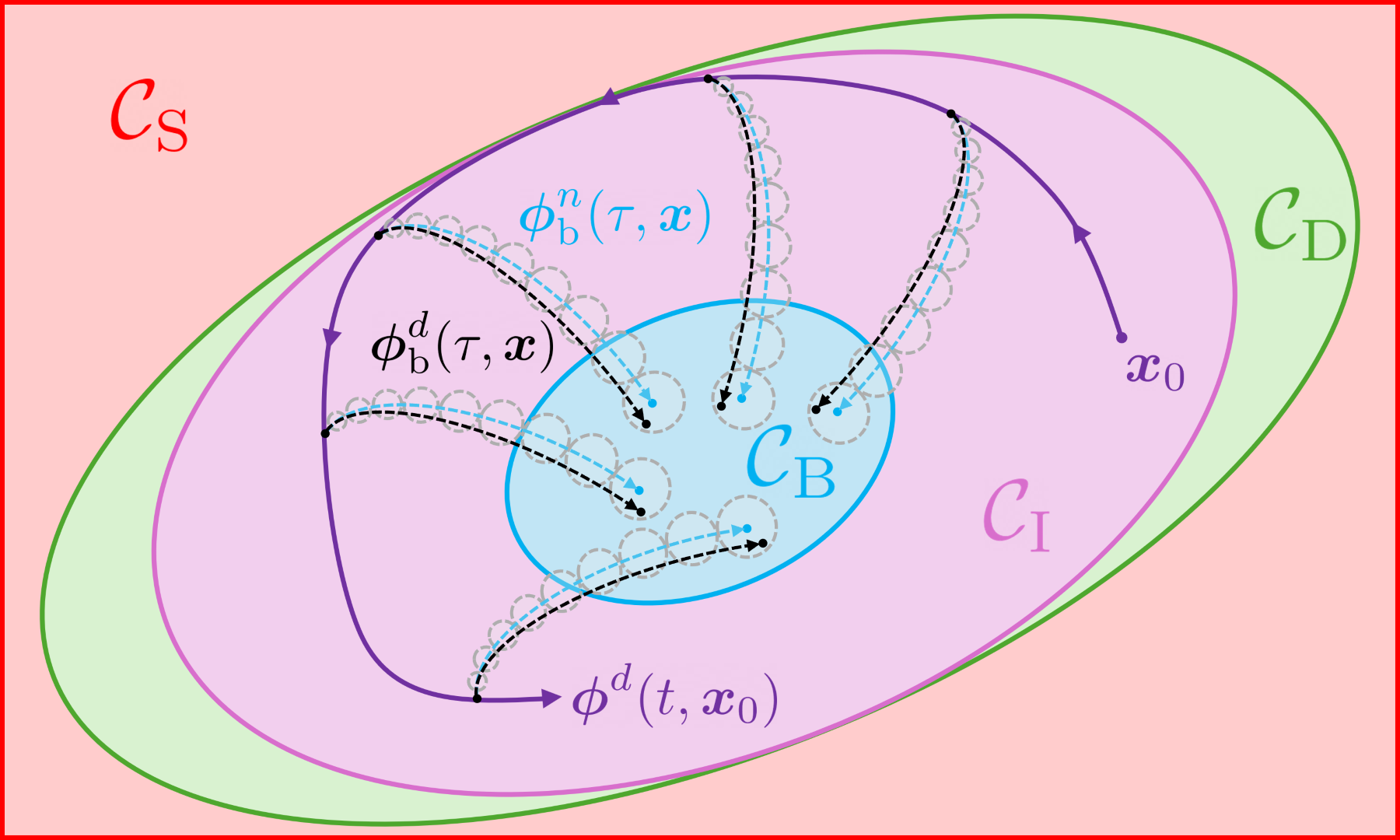}}
    % \centerline{\includegraphics[width=.88\columnwidth]{figs/evenwider.png}}
    \caption{Depiction of the proposed disturbance-robust safety-critical control framework. $\Ci$ represents a forward invariant subset of an unknown controlled invariant set $\Cd$ and guarantees safety of the disturbed system.}
    \label{fig: overview}
\end{figure}

\revised{The main contribution of this work is a novel approach to address controlled invariance and dynamics disturbances simultaneously through the formulation of \textit{disturbance-robust backup CBFs}. Unlike existing works, a controlled invariant set describing safety is not assumed to be known a priori, but is instead constructed online. We first derive forward invariance conditions for a subset inside a controlled invariant set of the disturbed system (displayed in Figure~\ref{fig: overview}). Then we robustify these conditions and integrate them with existing controllers via a quadratic program. The proposed framework guarantees safety for a broad class of nonlinear systems with limited control authority even in the presence of unknown, bounded disturbances. We demonstrate the effectiveness of the approach using two numerical simulations: an illustrative double integrator system and a spacecraft rotation example.}

\section{Preliminaries}

\subsection{Control Barrier Functions}

Consider a nonlinear control affine system of the form
\begin{align} \label{eq:affine-dynamics}
    \dot{\boldsymbol{x}} = f(\boldsymbol{x}) + g(\boldsymbol{x})\boldsymbol{u}, \nspace{8}
    \revised{\boldsymbol{x} \in \mathcal{X} \subseteq \mathbb{R}^n, \nspace{8}
    \boldsymbol{u} \in \mathcal{U} \subseteq \mathbb{R}^m,}
\end{align}
where $f:\mathcal{X} \rightarrow \mathbb{R}^n$ and $g:\mathcal{X} \rightarrow \mathbb{R}^{n \times m}$ are Lipschitz continuous functions. \revised{It is assumed that $\mathcal{U}$ is an $m$-dimensional convex polytope.}
For an initial condition $\boldsymbol{x}(0) = \boldsymbol{x}_0 \in \mathcal{X}$ if $\boldsymbol{u}$ is given by a locally Lipschitz feedback controller $k:\mathcal{X} \rightarrow \mathcal{U}$, $\boldsymbol{u}=k(\boldsymbol{x})$, the closed-loop system \eqref{eq:affine-dynamics} has a unique solution $\boldsymbol{\phi}^n(t, \boldsymbol{x}_0)$ over an interval of existence.

In the context of this work, safety is defined 
\revised{by membership to set $\Cs$}.
Safe controllers are ones that render this safe set forward invariant. 
A set $\mathcal{C} \subset \mathbb{R}^n$ is \textit{forward invariant} along \eqref{eq:affine-dynamics} if $\boldsymbol{x}_0 \in \mathcal{C} \implies \boldsymbol{\phi}^n(t, \boldsymbol{x}_0) \in \mathcal{C},$ for all $t > 0$.
Now, consider the safe set $\Cs$ as the 0-superlevel set of a continuously differentiable function $h : \mathcal{X} \rightarrow \mathbb{R}$ with $\Cs \triangleq \{\boldsymbol{x} \in \mathcal{X} : h(\boldsymbol{x}) \ge 0\}$,
% \begin{align} \label{eq:safeset}
%     \Cs \triangleq \{\boldsymbol{x} \in \mathcal{X} : h(\boldsymbol{x}) \ge 0\}, 
% \end{align}
where the gradient of $h$ along the boundary of $\Cs$ remains nonzero.
% , i.e., $h(\boldsymbol{x}) = 0 $ implies $ \nabla h(\boldsymbol{x}) \neq 0$.
\revised{
% CBFs are a powerful tool which connect forward invariance and safety.
A function $h : \mathcal{X} \rightarrow \mathbb{R}$ is a CBF \cite{ames_2017} for \eqref{eq:affine-dynamics} on $\Cs$ if there exists a class-$\mathcal{K}_{\infty}$ function\footnote{$\alpha : \mathbb{R}_{\ge 0} \rightarrow \mathbb{R}_{\ge 0}$ is a class-$\mathcal{K}_{\infty}$ function if it is continuous, $\alpha(0)=0$ and $\text{lim}_{x \rightarrow \infty} \nspace{2}\alpha(x) = \infty$.} $\alpha$ such that for all $\boldsymbol{x} \in \Cs$
\begin{equation*}
    \sup_{\boldsymbol{u} \in \mathcal{U}} \dot{h}(\boldsymbol{x},\boldsymbol{u}) \triangleq \underbrace{\nabla h(\boldsymbol{x}) f(\boldsymbol{x})}_{ L_f h(\boldsymbol{x})} + \underbrace{\nabla h(\boldsymbol{x}) g(\boldsymbol{x})}_{L_g h (\boldsymbol{x})}\boldsymbol{u} \ge  -\alpha(h(\boldsymbol{x})),
\end{equation*}
% where $L_f h$ and $L_g h$ are the Lie derivatives of $h$ along $f$ and $g$, respectively.
where $L_{(\cdot)} h$ is the Lie derivative of $h$ along function $(\cdot)$.
}
\begin{theorem}[\!\!\cite{ames_2017}] \label{thm: cbf}
If $h$ is a CBF for \eqref{eq:affine-dynamics} on $\Cs$, then any locally Lipschitz controller $k:\mathcal{X} \rightarrow \mathcal{U}$, $\boldsymbol{u}=k(\boldsymbol{x})$ satisfying 
\begin{align} \label{eq: cbf_condition}
    L_f h(\boldsymbol{x}) + L_g h (\boldsymbol{x}) \boldsymbol{u} \ge -\alpha(h(\boldsymbol{x}))
\end{align}
for all $\boldsymbol{x} \in \Cs$ renders the set $\Cs$ forward invariant.
\end{theorem}

% The sufficient condition for forward invariance of the safe set is linear in the control, often motivating the use of CBFs to supply safety constraints to a quadratic program (QP). 
For an arbitrary primary controller, $\boldsymbol{u}_{\rm p} \in \mathcal{U}$, it is possible to ensure the safety of \eqref{eq:affine-dynamics} by solving the following point-wise optimization problem for the safe control, $\boldsymbol{u}_{\rm safe}$:
\begin{align} 
% \begin{gathered} 
    \boldsymbol{u}_{\rm safe} = \underset{\boldsymbol{u} \in \mathcal{U}}{\text{argmin}} \mkern9mu & \frac{1}{2}\left\Vert \boldsymbol{u}_{\rm p}-\boldsymbol{u}\right\Vert^{2}
    % \quad \quad \quad \quad  (\text{CBF-QP})
    \tag{CBF-QP} \label{eq:cbf-qp} \\
    % \quad \quad \quad \quad
    \text{s.t.} \quad & L_f h(\boldsymbol{x}) + L_g h (\boldsymbol{x}) \boldsymbol{u} \ge -\alpha(h(\boldsymbol{x})). \nonumber
% \end{gathered}
\end{align}
A key challenge is obtaining an explicit representation of such a function $h$ where a safe control signal satisfying \eqref{eq: cbf_condition} can always be found. Depending on the safe set this may be difficult or impossible, especially for high dimensional systems. This therefore motivates the use of an extension of CBFs known as backup CBFs.

\subsection{Implicitly Defined Controlled Invariant Sets} \label{sec: baCBF}

First introduced in \cite{gurriet_online_2018} and expanded upon in \cite{gurriet_scalable_2020}, the backup CBF approach relies on \revised{obtaining} an implicitly defined controlled invariant set. \revised{A set $\mathcal{C} \subset \mathbb{R}^n$ is \textit{controlled invariant} if there exists a controller $k:\mathcal{X} \rightarrow \mathcal{U}$, $\boldsymbol{u}=k(\boldsymbol{x})$ which renders $\mathcal{C}$ forward invariant for \eqref{eq:affine-dynamics}.}

To construct an implicit controlled invariant set, first assume that we have defined a set $\Cs$ describing our state constraints, which is not necessarily controlled invariant. Now suppose that there exists a set within $\Cs$ which we call a {\em backup set}, $\Cb$, such that $\Cb \subset \Cs$. This set is defined similar to $\Cs$ with a continuously differentiable function $h_{\rm b}$, it is known to be controlled invariant, and it is made forward invariant by a {\em backup control law} defined by $\boldsymbol{u}_{\rm b}: \mathcal{X} \rightarrow \mathcal{U}$.
The closed-loop system under $\boldsymbol{u}_{\rm b}$ is denoted as
\begin{align} \label{eq: f_cl}
    f_{\rm cl}(\boldsymbol{x}) \triangleq f(\boldsymbol{x}) + g(\boldsymbol{x})\ub(\boldsymbol{x}).
\end{align}
 It is assumed that for any $\boldsymbol{x} \in \mathcal{X}$ there exists a unique solution $\boldsymbol{\phi}^{n}_{\rm b} : [0,T] \times \mathcal{X} \rightarrow \mathcal{X}$ which satisfies:
\begin{align} \label{eq: nomFlow}
    \skew{-3}\dot{\boldsymbol{\phi}^{n}_{\rm b}}(\tau,\boldsymbol{x}) = f_{\rm cl}(\phinom), \sspace \phinb{0}{\boldsymbol{x}} = \boldsymbol{x}.
\end{align}
The solution is the \textit{flow} of the system over the interval $[0,T]$ for $T \in \mathbb{R}_{>0}$ starting at state $\boldsymbol{x}$ under the backup control law $\ub$.
\noindent To obtain an implicitly defined controlled invariant set, $\Cbi$, satisfying $\Cb \subseteq \Cbi \subseteq \Cs$, one must ensure that the trajectory of the system under $\ub(\boldsymbol{x})$ remains in $\Cs$ over a finite time horizon, and that the final point in the trajectory lies within the backup set $\Cb$.
Therefore $\Cbi$ is defined as
\begin{align}
    \Cbi \triangleq \left\{ \boldsymbol{x} \in \mathcal{X} \,\middle|\, 
    \begin{array}{c}
    h(\phinom) \geq 0, \forall \nspace{1} \tau \in [0,T], \\
    h_{\rm b}(\phinomT) \geq 0 \\
    \end{array}
    \right\}.
\end{align}
\revised{The sufficient condition for forward invariance of $\Cbi$, and thus safety with respect to $\Cs$, is then}
\begin{subequations}\label{eq: nom_bcs1}
\begin{align} 
    \nabla h(\phinom)\stmnom\dot{\boldsymbol{x}} &\ge - \alpha (h(\phinom)), \label{eq: htraj_nom}  \\ 
    \!\!\nabla h_{\rm b}(\phinomT) \stmnomT \dot{\boldsymbol{x}} &\ge - \alpha_{\rm b} (h_{\rm b}(\phinomT)), \label{eq: hb_nom}
\end{align}
\end{subequations}
for all ${\tau \in [0,T]}$ and class-$\mathcal{K}_{\infty}$ functions $\alpha$ and $\alpha_{\rm b}$. \revised{Here,} ${\dot{\boldsymbol{x}}= f(\boldsymbol{x}) + g(\boldsymbol{x})\boldsymbol{u}}$ and ${\stmnom \triangleq \partial \phinom/\partial \boldsymbol{x}}$ is the sensitivity matrix, or state-transition matrix (STM), which captures the sensitivity of the flow to perturbations in the initial condition $\boldsymbol{x}$. The STM is the solution to
% the differential equation
\begin{equation}
  \begin{gathered} 
    {\dot{\boldsymbol{\Phi}}}_{\rm b}^{n}(\tau, \boldsymbol{x}) = \jac(\phinom)\stmnom,
    \sspace
    \boldsymbol{\Phi}_{\rm b}^{n}(0,\boldsymbol{x}) = \revised{\boldsymbol{I}},
\end{gathered}
\end{equation}
where $\jac$ is the Jacobian of the closed-loop backup dynamics \eqref{eq: f_cl} evaluated at $\phinom$ and $\revised{\boldsymbol{I}}$ is the $n \times n$ identity matrix. 

\revised{Because the inequality in \eqref{eq: htraj_nom} represents an infinite number of constraints, in practice these are discretized and enforced at discrete times along the flow.}
To ensure safety between sample points, \eqref{eq: htraj_nom} is tightened via a constant $\varepsilon_{\Delta}$\revised{\cite[Thm.~3]{gurriet_thesis}}:
\begin{align}
    \varepsilon_{\Delta} \geq \frac{\Delta}{2} \mathcal{L}_{h} \sup_{\boldsymbol{x} \in \Cs} \norm{f_{\rm cl}(\boldsymbol{x})},
\end{align}
where $\Delta \in \mathbb{R}_{>0}$ is a discretization time step satisfying  $T/\Delta \in \mathbb{N}$, $\mathcal{L}_{h} \in \mathbb{R}_{>0}$ is the Lipschitz constant of $h$ with respect to the Euclidean norm and $\sup_{\boldsymbol{x} \in \Cs} \norm{f_{\rm cl}(\boldsymbol{x})}$ is the maximal velocity of the backup vector field. 

\revised{As in
% \hyperref[eq:cbf-qp]{(CBF-QP)}
\eqref{eq:cbf-qp}, the safety of \eqref{eq:affine-dynamics} can be enforced for a primary controller, $\boldsymbol{u}_{\rm p} \in \mathcal{U}$, by solving an optimization problem for the safe control with constraints \eqref{eq: nom_bcs1}, where the right-hand side of \eqref{eq: htraj_nom} is replaced by $- \alpha (h(\phinom) - \varepsilon_{\Delta})$.}

\section{Disturbance Robustness}

While the \revised{standard backup set method reviewed in \Cref{sec: baCBF}} can guarantee safety for a system in which the dynamics are perfectly known, in practice there are always unmodeled parameters or external disturbances which perturb the dynamics. Therefore, it is desirable to leverage the advantages offered by the backup CBF approach, for dynamics with process disturbances. As such, consider the system
\begin{align} \label{eq: disturbed_dyn}
    \dot{\boldsymbol{x}} = f(\boldsymbol{x}) + g(\boldsymbol{x})\boldsymbol{u} + \boldsymbol{d}_x,
\end{align}
where ${\boldsymbol{d}_x \in \mathcal{D}_x \subseteq \mathbb{R}^n}$ is an unknown additive process disturbance and there exists a constant $\xi \in \mathbb{R}_{>0}$ such that ${\norm{\boldsymbol{d}_x} \leq \xi}$. For an initial condition $\boldsymbol{x}(0) = \boldsymbol{x}_0 \in \mathcal{X}$ and a locally Lipschitz controller $\boldsymbol{u}=k(\boldsymbol{x})$, if $\boldsymbol{d}_x$ is piecewise continuous in time, the closed-loop system \eqref{eq: disturbed_dyn} has a unique solution $\phidg{t}{\xzero}$ over an interval of existence.

We assume that a backup control law $\ub$ can be obtained which renders a backup set $\Cb$ inside $\Cs$ robustly forward invariant. This is made more precise below.
\begin{assumption} \label{assump: robust_inv}
     The backup controller $\boldsymbol{u}_{\rm b}$ renders the backup set $\Cb$ forward invariant along \eqref{eq: disturbed_dyn} for any disturbance $\boldsymbol{d}_x$ which satisfies $\norm{\boldsymbol{d}_x} \leq \xi$.
\end{assumption}
Backup sets are often defined by a level set of a quadratic Lyapunov function based on the linearized dynamics about a stabilizable equilibrium point \cite{gurriet_scalable_2020,chen2021backupcontrolbarrierfunctions}, and a simple feedback controller such as a linear quadratic regulator can be used to render this set forward invariant. 
Techniques to robustify such quadratic Lyapunov functions have been studied in the literature \cite{jankovic_robust_2018},\cite[Ch. 13.1]{khalil2002nonlinear},\cite[Ch. 3]{freeman_robust_1996}. 
While this robustification may result in a smaller backup set, this set is expanded to generate a larger controlled invariant set.

Next we define two separate flows: the nominal and the disturbed backup flow. The nominal backup flow $\phinom$ satisfies \eqref{eq: nomFlow} under the robust control law $\ub$, while the disturbed backup flow, denoted $\phid$, is the solution to
\begin{align} \label{eq: x_ddot}
    \skew{-1}\dispdot{\boldsymbol{\phi}^{d}_{\rm b}}(\tau,\boldsymbol{x}) = f_{\rm cl}(\phid) + \boldsymbol{d}_x, \sspace \boldsymbol{\phi}^{d}_{\rm b}(0,\boldsymbol{x}) = \boldsymbol{x}.
\end{align}
Again, it is assumed that for any $\boldsymbol{x} \in \mathcal{X}$ there exists a unique solution $\boldsymbol{\phi}^{d}_{\rm b} : [0,T] \times \mathcal{X} \rightarrow \mathcal{X}$ to \eqref{eq: x_ddot}. Consider $\Cd \subseteq \Cs$
\begin{align} \label{eq: Cd}
    \hspace{-.09cm}
    \Cd \triangleq \left\{ \boldsymbol{x} \in \mathcal{X} \,\middle|\, 
    \begin{array}{c}
    h(\phid) \geq 0, \forall \nspace{1} \tau \in [0,T], \\
    h_{\rm b}(\phidT) \geq 0 \\
    \end{array}
    \right\}.
\end{align}
We are interested in forward invariance conditions for $\Cd$, but since the disturbance is unknown, we will instead derive forward invariance conditions for a set which over-approximates the disturbed flow. Consider a new set, $\Ci$, defined by 
\begin{align} \label{eq: Ci}
    \hspace{-.09cm}
    \Ci \triangleq \left\{ \boldsymbol{x} \in \mathcal{X} \,\middle|\, 
    \begin{array}{c}
    h(\phinom) \geq \epsilon_\tau, \forall \nspace{1} \tau \in [0,T], \\
    h_{\rm b}(\phinomT) \geq \epsilon_{\rm b}  \\
    \end{array}
    \right\}.
\end{align}
The set is entirely governed by the nominal trajectory and additional tightening terms $\epsilon_{\tb}$ and $\epsilon_{\rm b}$. For judiciously chosen values of $\epsilon_{\tb}$ and $\epsilon_{\rm b}$, we show that $\Ci$ is a subset of $\Cd$.
\begin{lemma} \label{lemma: ci_subset}
Let $\mathcal{L}_{h}$ and $\mathcal{L}_{h_{\rm b}}$ be the Lipschitz constants of $h$ and $h_{\rm b}$, respectively, and let $\delta_{{\rm max}}(\tb)$ be a norm bound on the deviation between $\phinom$ and $\phid$ at time $\tbSet$:
\begin{equation}
    \norm{\phinom - \phid} \leq \delta_{{\rm max}}(\tb),
\end{equation}
for all $\boldsymbol{x} \in \Cs$.
If $\epsilon_{\tb} \geq \mathcal{L}_{h} \delta_{{\rm max}}(\tb)$ holds for all $\tbSet$ and $\epsilon_{\rm b} \geq \mathcal{L}_{h_{\rm b}} \delta_{{\rm max}}(T)$ also holds, then $\Ci \subseteq \Cd$.
\begin{proof}
Consider any state $\boldsymbol{x} \in \Ci$. Membership to $\Ci$ implies that $h(\phinom) \geq \epsilon_{\tb} \geq \mathcal{L}_{h} \delta_{{\rm max}}(\tb)$.
Hence it follows that
\begin{align*}
     h(\phid) & = h(\phinom) - \big( h(\phinom) - h(\phid) \big) \\
     & \geq \mathcal{L}_{h} \delta_{{\rm max}}(\tb) - \big| h(\phinom) - h(\phid) \big|.
\end{align*}
By Lipschitz continuity of the constraint function $h$
\begin{align*}
    & | h (\phinom) - h (\phid) | \\
    & \quad \leq \mathcal{L}_{h} \norm{\phinom - \phid} \leq \mathcal{L}_{h} \delta_{{\rm max}}(\tb),
\end{align*}
we obtain ${h(\phid) \geq 0}$ for any $\boldsymbol{x} \in \Ci$.
Similar logic can be applied to the constraint on the reachability of the backup set. For any $\boldsymbol{x} \in \Ci$, the nominal backup trajectory from $\boldsymbol{x}$ satisfies $h_{\rm b}(\phinomT) \geq \epsilon_{\rm b}\geq \mathcal{L}_{h_{\rm b}} \delta_{{\rm max}}(T)$, and we have
\begin{align*}
    | h_{\rm b} (\phinomT) - h_{\rm b}(\phidT) | \leq \mathcal{L}_{h_{\rm b}} \delta_{{\rm max}}(T).
\end{align*}
These guarantee that $h_{\rm b}(\phidT) \geq 0$.
Thus, all the functions which define $\Cd$ are nonnegative, meaning that $\boldsymbol{x} \in \Cd$ for all $\boldsymbol{x} \in \Ci$, and so $\Ci \subseteq \Cd$.
\end{proof}
\end{lemma}
\noindent \Cref{lemma: ci_subset} assumes that a time-varying bound on the deviation between the nominal and disturbed backup flow, $\delta_{\rm max}(\tau)$, can be found. While problem-specific bounds can be obtained, we utilize a generalization of the Gronwall-Bellman inequality to obtain a bound for a wide class of nonlinear systems.
\begin{lemma}[Theorem 2.5 in \cite{khalil2002nonlinear}] \label{lemma: GW}
For systems \eqref{eq: nomFlow} and \eqref{eq: x_ddot}, let $f_{\rm cl}$ be locally Lipschitz on $\mathcal{X}$ with Lipschitz constant $\mathcal{L}_{\rm cl}$ and $\boldsymbol{d}_x$ be piecewise continuous in $\tau$ on $[0, T]$. If $\norm{\boldsymbol{d}_x} \leq \xi$ for all $\boldsymbol{d}_x$ and some $\xi > 0$, then for all $\tau \in [0,T]$ one has
\begin{align*}
    \norm{\phinom - \phid} \leq \frac{\xi}{\mathcal{L}_{\rm cl}}(e^{\mathcal{L}_{\rm cl}\tau} - 1) \triangleq \delta_{{\rm max}}(\tb).
\end{align*}
% for all $\tau \in [0,T]$.
\end{lemma}
\begin{remark} \label{rem: boundsContr}
    \revised{Backup strategies often drive the system to an equilibrium and thus may be (at least weakly) contracting.}
    When contraction bounds on the deviation between the disturbed and nominal backup flow can be obtained, $\delta_{{\rm max}}(\tb)$ can be made less conservative, 
    \revised{and it will converge to a near-constant value as $T$ increases.}
    Details on such bounds can be found in \cite[Corollary 3.17]{contraction_bullo}. A contraction bound is used effectively in the spacecraft rotation example in \Cref{sec: spacecraft}. For linear systems, flow deviation bounds can be even tighter.
\end{remark}

Using the definition of $\Cd$ and the corresponding robust backup controller $\ub$, we now examine the properties of $\Cd$.
\begin{lemma} \label{lemma: cd_inv}
The set $\Cd$ is controlled invariant, and the robust backup controller $\ub$ renders $\Cd$ forward invariant along \eqref{eq: disturbed_dyn}, such that 
\begin{align}
    \boldsymbol{x} \in \Cd \implies \boldsymbol{\phi}_{\rm b}^{d} (\vartheta, \boldsymbol{x}) \in \Cd, \forall \nspace{1} \vartheta \geq 0.
\end{align}
\begin{proof}
    From the definition of $\Cd$ and with \Cref{assump: robust_inv}
    \begin{align} \label{eq: l1_1}
        \boldsymbol{x} \in \Cd \implies \boldsymbol{\phi}_{\rm b}^{d} (\tau, \boldsymbol{x}) \in \Cb \subseteq \Cs, \forall \nspace{1} \tau \geq T.
    \end{align}
    By definition, the flow is recursive in nature and thus for any $\boldsymbol{x} \in \mathbb{R}^n$ and $\tau, \vartheta \geq 0$, $\boldsymbol{\phi}_{\rm b}^{d} (\tau + \vartheta, \boldsymbol{x}) = \boldsymbol{\phi}_{\rm b}^{d} (\tau, \boldsymbol{\phi}_{\rm b}^{d} (\vartheta, \boldsymbol{x}))$.
    Using \eqref{eq: l1_1} and the recursive property of the flow 
    \begin{align} \label{eq: l1_Tb}
        \boldsymbol{x} \in \Cd \implies \boldsymbol{\phi}_{\rm b}^{d} (T, \boldsymbol{\phi}_{\rm b}^{d} (\vartheta, \boldsymbol{x})) \in \Cb, \forall \nspace{1}\vartheta \geq 0.
    \end{align}
    From \eqref{eq: l1_1} and by definition \eqref{eq: Cd} $\boldsymbol{x} \in \Cd \implies \boldsymbol{\phi}_{\rm b}^{d} (\tau, \boldsymbol{x}) \in \Cs, \forall \nspace{1} \tau \geq 0$.
    Using the recursive property once more
    \begin{align} \label{eq: l1_done}
        \nspace{-3} \boldsymbol{x} \in \Cd \nspace{-6} \implies \nspace{-6}\boldsymbol{\phi}_{\rm b}^{d} (\tau, \boldsymbol{\phi}_{\rm b}^{d} (\vartheta, \boldsymbol{x})) \nspace{-2} \in \nspace{-2} \Cs, \forall \nspace{1} \tau \nspace{-1}\in \nspace{-1}[0, T], \forall \vartheta \nspace{-2}\geq \nspace{-2}0.
    \end{align}
    Definition \eqref{eq: Cd} with \eqref{eq: l1_Tb} and \eqref{eq: l1_done} completes the proof.
\end{proof}
\end{lemma}

While the controlled invariance of $\Cd$ has been established, the conditions on $\boldsymbol{u}$ for forward invariance cannot yet be obtained as $\Cd$ itself is unknown. This motivates the following theorems. 
\begin{theorem} \label{thm: controlInv_cd}
For any $\boldsymbol{x} \in \Ci$, there exists a controller $\boldsymbol{u}$ such that $\boldsymbol{\phi}^{d} (\vartheta, \boldsymbol{x}) \in \Cd \subseteq \Cs, \forall \nspace{2} \vartheta \geq 0$. 
\begin{proof}
By \Cref{lemma: ci_subset}, ${\boldsymbol{x} \in \Ci \nspace{-6} \implies \boldsymbol{x} \in \Cd}$, and by \Cref{lemma: cd_inv}, $\boldsymbol{u}_{\rm b}$ ensures ${\boldsymbol{\phi}^{d} (\vartheta, \boldsymbol{x}) \in \Cd \subseteq \Cs}$, ${\forall \nspace{2} \vartheta \geq 0}$.
\end{proof}
\end{theorem}
We are now ready to establish the conditions that enable a controller to ensure the robust safety of \eqref{eq: disturbed_dyn}. From the definition of $\Ci$ we have
 \begin{align}
    \begin{gathered}
    \dot{h}(\phinom, \boldsymbol{u}) \ge - \alpha(h(\phinom) - \epsilon_\tau), \\
    \dot{h}_{\rm b}(\phinomT, \boldsymbol{u}) \ge -\alpha_{\rm b}(h_{\rm b}(\phinomT) - \epsilon_{\rm b}),
    \end{gathered}
\end{align}
where by expanding the total derivatives for system \eqref{eq: disturbed_dyn} this becomes, $\forall \tau \in [0,T],$
\begin{align} \label{eq: nagumo_ci}
    \begin{gathered}
    \nspace{-4}\nabla h(\phinom) \stmnom \dot{\boldsymbol{x}}^d \ge - \alpha(h(\phinom) - \epsilon_\tau), \\ 
    \nabla h_{\rm b}(\phinomT) \stmnomT \dot{\boldsymbol{x}}^d \ge -\alpha_{\rm b}(h_{\rm b}(\phinomT) - \epsilon_{\rm b}).
    \end{gathered}
\end{align}
Here, $\dot{\boldsymbol{x}}^d \triangleq f(\boldsymbol{x}) + g(\boldsymbol{x})\boldsymbol{u} + \boldsymbol{d}_x$. Using this expansion, we can show that a \revised{controller} which realizes forward invariance of $\Ci$ keeps the disturbed system safe, and that conditions for such a controller can be directly computed, despite the unknown disturbance.

\begin{theorem} \label{thm: mainResult}
    If any controller $\boldsymbol{u}$ satisfies 
    \begin{align} \label{eq: mainthmConstraints}
        \begin{gathered}
        \nabla h(\phinom) \stmnom \big( f(\boldsymbol{x}) + g(\boldsymbol{x}) \boldsymbol{u} \big) - \eta \ge \\ 
        \hspace{2cm} - \alpha (h(\phinom) - \epsilon_{\tau}), \\ 
        \nabla h_{\rm b}(\phinomT) \stmnomT \big( f(\boldsymbol{x}) + g(\boldsymbol{x}) \boldsymbol{u} \big) - \eta_{\rm b} \ge \\
        \hspace{2.4cm} - \alpha_{\rm b} (h_{\rm b}(\phinomT) - \epsilon_{\rm b}),
        \end{gathered}
    \end{align}
    with robustness terms defined by 
    \begin{align*}
    \begin{gathered}
        \eta \triangleq \xi \norm{\nabla h(\phinom) \stmnom}, \\
        \eta_{\rm b} \triangleq \xi \norm{\nabla h_{\rm b}(\phinomT) \stmnomT},
    \end{gathered}
    \end{align*}
    then $\boldsymbol{x}_0 \in \Ci \implies \phidg{t}{\xzero} \in \Ci \subseteq \Cd \subseteq \Cs, $ for all $ \nspace{1} t > 0$.
\begin{proof}
    As done in \cite{jankovic_robust_2018}, the robustness terms $\eta$ and $\eta_{\rm b}$ upper-bound the unknown $\boldsymbol{d}_x$ term in~\eqref{eq: nagumo_ci}.
    Thus the condition~\eqref{eq: mainthmConstraints} implies~\eqref{eq: nagumo_ci}.
    From a direct application of Theorem~\ref{thm: cbf} to system \eqref{eq: disturbed_dyn}, we obtain that \eqref{eq: nagumo_ci}
    ensures ${\phidg{t}{\xzero} \in \Ci}$, ${\forall t > 0}$ for any ${\boldsymbol{x}_0 \in \Ci}$.
    From \Cref{lemma: ci_subset}, we have $\Ci \subseteq \Cd$.
\end{proof}
\end{theorem}
Naturally, the original backup CBF constraints \eqref{eq: nom_bcs1} are recovered in the absence of disturbances (i.e., $\xi = 0$). As the constraints in \eqref{eq: mainthmConstraints} are continuous in $\tau$, the trajectory is again discretized \revised{similar to \cite[Thm. 3]{gurriet_thesis}} and appropriately tightened via a constant term $\varepsilon_{\Delta}$ where
\begin{align}
    \varepsilon_{\Delta} \geq \frac{\Delta}{2} \mathcal{L}_{h} (\sup_{\boldsymbol{x}\in \Cs} \left \Vert f_{\rm cl}(\boldsymbol{x}) \right \Vert + \xi).
\end{align}

The result of \Cref{thm: mainResult} is now ready to be directly utilized in a new point-wise optimal controller accounting for disturbances. The \textit{Disturbance-Robust Backup CBF (DR-bCBF)} optimization problem is written as:
\begin{align*} 
% \begin{gathered} 
    \boldsymbol{u}_{\rm safe} & = \underset{\boldsymbol{u} \in \mathcal{U}}{\text{argmin}} \mkern9mu \frac{1}{2}\left\Vert \boldsymbol{u}_{\rm p}-\boldsymbol{u}\right\Vert^{2} \quad \tag{DR-bCBF-QP} \label{eq:dr-bacbf-qp} \\
    % \quad \quad \quad  (\text{DR-bCBF-QP}) \\ 
    \text{s.t.  } &
    \nabla h(\phinom) \stmnom  \big( f(\boldsymbol{x}) + g(\boldsymbol{x}) \boldsymbol{u} \big) - \eta \ge \\
    & \nquad{8} - \alpha (h(\phinom) - \epsilon_{\tb} - \varepsilon_{\Delta}), \\ 
    & \nabla h_{\rm b}(\phinomT) \stmnomT \big( f(\boldsymbol{x}) + g(\boldsymbol{x}) \boldsymbol{u} \big) - \eta_{\rm b} \ge \\
    & \nquad{9} - \alpha_{\rm b} (h_{\rm b}(\phinomT) - \epsilon_{\rm b}),
% \end{gathered}
% \label{eq:dr-bacbf-qp}
\end{align*}
for all $\tau \in \{0, \Delta, \dots, T \}$. As before, $\Delta \in \mathbb{R}_{>0}$ is a discretization time step satisfying  $T/\Delta \in \mathbb{N}$. \revised{Because $\epsilon_{\tau}$ and $\epsilon_{\rm b}$ only depend on a priori known values and $\tau$, they can be pre-computed and reused each time
% \hyperref[eq:dr-bacbf-qp]{(DR-bCBF-QP)}
\eqref{eq:dr-bacbf-qp} is solved.
Furthermore, the robustness terms $\eta$ and $\eta_{\rm b}$ depend on values that must be computed for the standard backup set method already, hence disturbance robustness adds negligible computational cost.}
\begin{remark}
    From \Cref{thm: controlInv_cd}, $\Cd$ is controlled invariant, however the controlled invariance of $\Ci$ itself cannot be proven without additional assumptions on $\boldsymbol{u}_{\rm b}$ and $\Cb$. Therefore, the feasibility of the optimization problem
    % \hyperref[eq:dr-bacbf-qp]{(DR-bCBF-QP)}
    \eqref{eq:dr-bacbf-qp} is not guaranteed. However, in the case that the optimization problem becomes infeasible, the robust backup control law $\boldsymbol{u}_{\rm b}$ can be used to stay in $\Cd$ until the optimization problem becomes feasible again. 
    % \revised{Thus a safe control signal satisfying the input bounds can always be found.}
\end{remark}

\section{Numerical Examples}
\revised{
In this section we demonstrate the effectiveness of the proposed method in assuring safety under bounded disturbances using two simulation examples.
Code and videos are available at: \href{https://github.com/davidvwijk/DR-bCBF}{https://github.com/davidvwijk/DR-bCBF}}.

\subsection{Double Integrator}

Consider a simple example of a double integrator given by 
\begin{align}
    \dot{\boldsymbol{x}} = 
    \revised{
    \begin{bmatrix}
        x_2,
        u
    \end{bmatrix}^{T}}
    \nspace{-4}
    + \boldsymbol{d}_x,
\end{align}
with a state vector $\boldsymbol{x} = [x_1, x_2]^{T} \in \mathbb{R}^2$ where $x_1$ is the position and $x_2$ is the velocity, and an acceleration control variable ${u \in \mathcal{U} = [-1, 1]}$. The safe set is defined as ${\Cs \triangleq \{\boldsymbol{x} \in \mathbb{R}^2 : -x_1 \geq 0 \}}$. The unknown additive process disturbance is bounded with $\norm{\boldsymbol{d}_x} \leq \xi \in \mathbb{R}_{>0}$. In this particular example, the unknown disturbance is constant, with ${\boldsymbol{d}_x = \xi \frac{\boldsymbol{v}}{\norm{\boldsymbol{v}}}}$ where ${\boldsymbol{v} = [1,1]^{T}}$, and ${\xi = 0.08}$. The backup control law ${\boldsymbol{u}_{\rm b}(\boldsymbol{x}) = -1}$ brings the system to the backup set ${\Cb \triangleq \{\boldsymbol{x} \in \mathbb{R}^2 : -x_1 \geq 0, -x_2 \geq 0 \}}$ as long as $\xi < 1$. The discretization time step for computing the nominal backup flow is ${\Delta = 0.02 \nspace{4}{\rm s}}$. The primary control law is ${\boldsymbol{u}_{\rm p} = 1}$ which drives the system to the right half-plane (unsafe region).

\begin{figure} %[ht!]
    \centering
    \centerline{\includegraphics[width=.95\columnwidth]{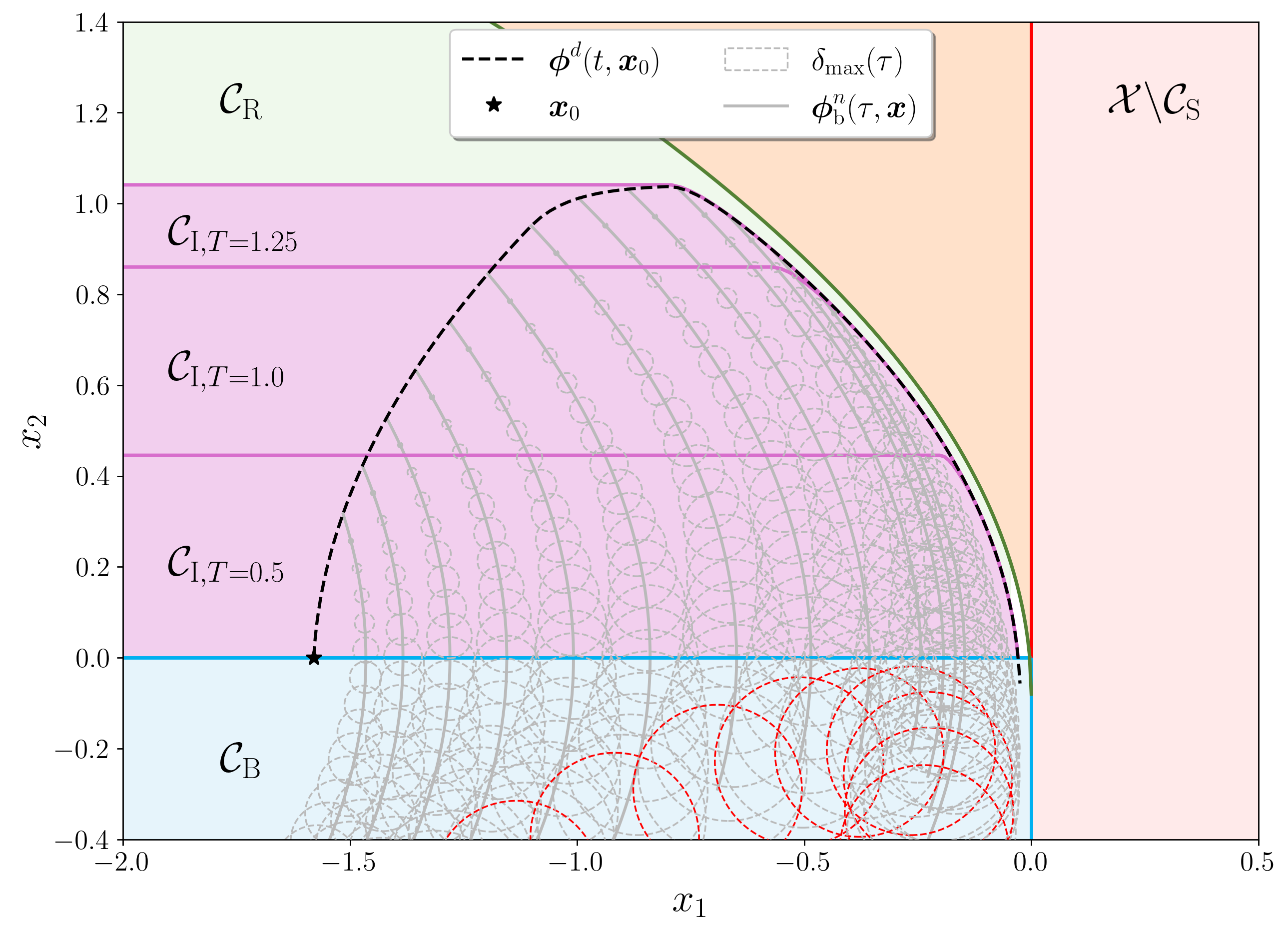}}
    \vspace{-.2cm}
    \caption{Phase space visualization of safety-critical control for a double integrator system under bounded disturbances, using the proposed disturbance-robust backup control barrier function approach. \revised{Nominal backup trajectories in gray emanate from the disturbed trajectory (dotted black line) and the gray circles centered on the nominal trajectories are Gronwall norm balls from \Cref{lemma: GW}. The Gronwall norm ball at $\tau = T$, colored in red, is always contained in $\Cb$, as required by \eqref{eq: Ci}.}}
    \label{fig: phaseSpace_dbint}
\end{figure}

Using the proposed disturbance-robust backup CBF approach, the forward invariant set $\Ci$ is computed for various values of $T$ between $0.5 \nspace{4}{\rm s}$ and $1.25 \nspace{4}{\rm s}$, plotted on \Cref{fig: phaseSpace_dbint} in pink. The backup set, $\Cb$, is plotted in blue and a robust controlled invariant set $\mathcal{C}_{\rm R}$ is plotted in green. For this simple linear system, such a set can be computed by analytically solving for the flow and accounting for worst-case disturbances. The black dotted line represents the trajectory of the disturbed system using the proposed controller, with an integration horizon $T = 1.25 \nspace{4}{\rm s}$.
Notably, safety is maintained along this trajectory, as stated by Theorem~\ref{thm: mainResult}.

The plots show that as the backup horizon $T$ is increased, the forward invariant set $\Ci$ increases in size, up to a certain point. Since the Gronwall bound grows exponentially with time, the longer the backup horizon is, the larger the final bound, $\delta_{\rm max}(T)$, will be. If $T$ is too large, the constraint on the terminal point of the nominal backup trajectory will dominate, and the set $\Ci$ will begin to shrink. This therefore introduces a trade-off as $T$ cannot be made arbitrarily large when using \Cref{lemma: GW}. Naturally, as the disturbance bound $\xi$ increases, the size of $\Ci$ will shrink since the bound on the backup flows is proportional to $\xi$.

\subsection{Rigid Body Spacecraft Rotation} \label{sec: spacecraft}

\begin{figure*}
\vspace{-.5cm}
\setlength{\linewidth}{\textwidth}
\setlength{\hsize}{\textwidth}
\centering
{\includegraphics[width=0.424\textwidth]{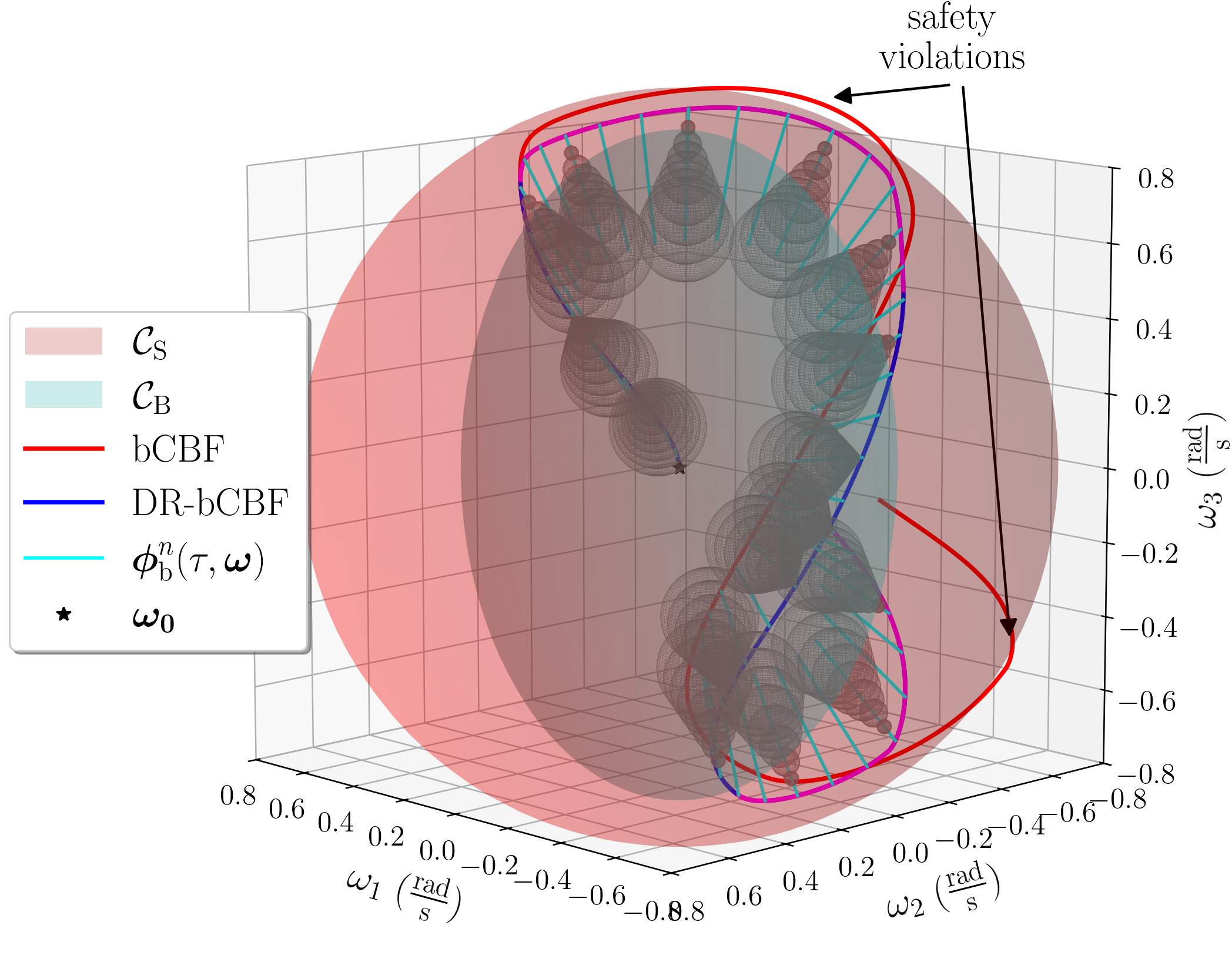} \label{subfig: 3d}}
\hspace{0.006\textwidth}
{\includegraphics[width=0.54\textwidth]{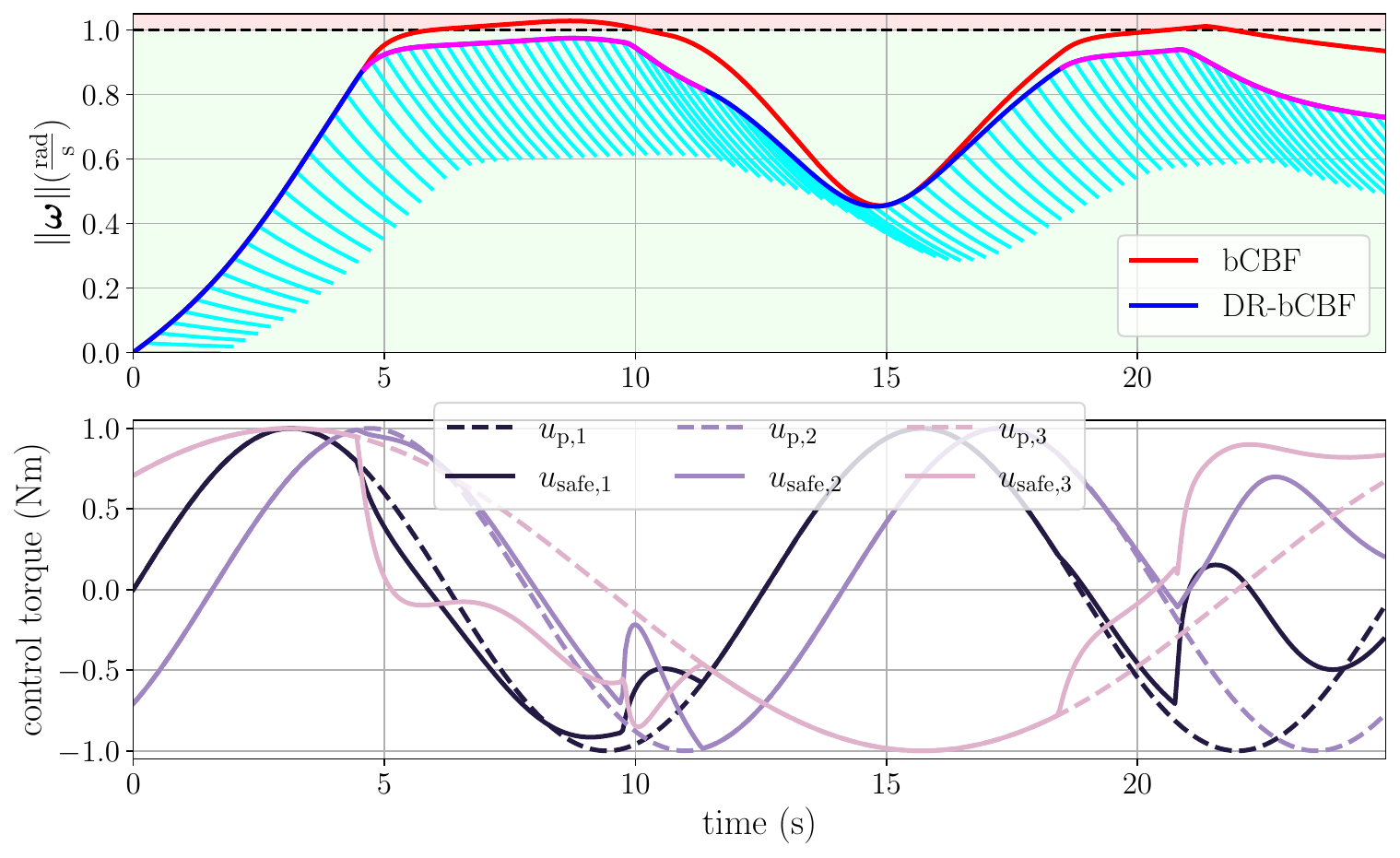}\label{subfig: norm_u}}
\vspace{-.3cm}
\caption{\revised{Simulation results for a rigid body spacecraft, comparing the proposed disturbance-robust backup CBF approach and the standard backup CBF formulation.}
(\textbf{Left}) State-space visualization of angular velocity components showing the trajectory of the angular velocity vector over time. \revised{The objective is to keep the trajectory within the red sphere (safe region). The standard approach violates safety due to the disturbance, while the proposed disturbance-robust method does not. Magenta sections of the blue trajectory indicate that the primary control signal, $\boldsymbol{u}_{\rm p}$, has been modified to assure safety. Wire-frame spheres represent the contraction norm balls along the nominal backup flow in cyan.}
(\textbf{Right}) Angular velocity norm over time \revised{for both approaches} (\textbf{top}), and commanded primary control and actual (safe) control signal over time \revised{for the robust approach} (\textbf{bottom}).}
\label{fig: spacecraftRot} 
\end{figure*}

Consider next an example of a rigid body spacecraft with a known inertia tensor in the body frame given by $\boldsymbol{J}$, where the dynamics of the angular velocities can be described by Euler's rotational equations of motion
\begin{align}
\label{eq:eulersEOM}
    \dot{\boldsymbol{{\omega}}} = \boldsymbol{J}^{-1}(-\w\times\boldsymbol{J}\boldsymbol{\omega} + \boldsymbol{u}) + \boldsymbol{d}_x.
\end{align} 
Here, $\boldsymbol{u} \in \mathcal{U} = [-1,1]^3 \nspace{4}{\rm Nm}$ is the control torque vector that can be applied by the spacecraft to control the angular velocity and $\boldsymbol{d}_x$ is an unknown but bounded additive disturbance vector, such that $\norm{\boldsymbol{d}_x} \leq \xi \in \mathbb{R}_{>0}$. The states of interest are the angular velocity vector elements in the body frame.

The safety objective is to ensure that the norm of the angular velocity vector of the spacecraft does not exceed a maximum value, to prevent damage to onboard sensors. The safe set is therefore $\mathcal{C}_{\rm S} = \{\boldsymbol{\omega} \in \mathbb{R}^{3} : h(\boldsymbol{\omega}) \ge 0\}$ where $h(\boldsymbol{\omega}) = \omega_{\rm max}^2 - \left \Vert \boldsymbol{\omega} \right \Vert^2 \ge 0$ and $\omega_{\rm max} \in \mathbb{R}_{>0}$ represents the maximum allowable angular velocity. The primary controller is given by 
${\boldsymbol{u}_{\rm p}(t) = \text{sin}([\frac{t}{2}, \frac{t}{2} - \frac{\pi}{4}, \frac{t}{4} + \frac{\pi}{4}]^{T})}$. The robust backup control law $ \ub(\w) = -k_b\J\w + \w \times \J \w$ renders the backup set $\Cb \triangleq \{ \w \in \mathbb{R}^{3} :  h_{\rm b}(\w) = \gamma - \frac{1}{2}\w^{T}\J\w \ge 0\}$ robustly forward invariant for sufficiently large gain $k_b$. $\Cb$ is a level set of the spacecraft's rotational energy, defined by the scalar $\gamma$. \revised{Any} $k_b > (\lambda_{\rm max} \xi) / ( \sqrt{2\gamma \lambda_{\rm min}})$ ensures $\dot{h}_{\rm b}(\w,\boldsymbol{u}) \ge 0$ at the boundary of the level set for \eqref{eq:eulersEOM}, where $\lambda_{\rm max}$ and $\lambda_{\rm min}$ are the maximum and minimum eigenvalues of $\J$, respectively. The proof is omitted for brevity.

It is straightforward to verify that the closed-loop nominal backup dynamics are strongly contracting with a rate of $k_b$ since $f_{\rm cl}(\w) = -k_b \w$. Because the log norm of the closed-loop Jacobian ($\partial f_{\rm cl}/\partial \w$) is upper-bounded by $-k_b$, the disturbed and nominal backup flows can be bounded as
\begin{align} \label{eq: contractionBnds}
    \norm{\phinb{\tau}{\w} - \phidb{\tau}{\w}} \leq \frac{\xi}{k_b}(1-e^{-k_b \tau}),
\end{align}
by \cite[Corollary 3.17]{contraction_bullo}. This yields tighter $\epsilon_\tau$ and $\epsilon_{\rm b}$ terms than the general Gronwall bound.

For the simulations, $\omega_{\rm max} = 1 \nspace{4} \rm rad/s$, $\gamma = 2 \rm  \nspace{4} J$, $\xi = 0.1 \nspace{4} {\rm rad/s^2}$ and \revised{$\J$ is diagonal with elements $[12,12,5] \nspace{4} {\rm kg \nspace{1}m^2}$.}
The discretization time step used for computing the nominal backup flow is $\Delta = 0.05 \nspace{4}{\rm s}$ and the integration horizon is $T = 1.75 \nspace{4}{\rm s}$. The disturbance vector is time-varying, given by $\boldsymbol{d}_x(t) = \xi\frac{\boldsymbol{v}(t) }{\| \boldsymbol{v}(t) \|}$ where  $\boldsymbol{v}(t) = \text{sin}([\frac{t}{2} + \frac{\pi}{2}, \frac{t}{2}, \frac{t}{2} - \frac{\pi}{2}]^{T})$. 

\Cref{fig: spacecraftRot} \revised{compares our disturbance-robust backup CBF method} using the contraction bounds in \eqref{eq: contractionBnds} \revised{with the standard backup CBF approach}.
\revised{Our approach obeys the norm constraint on the angular velocity in the presence of \revised{unknown} time-varying disturbances, while the standard backup CBF approach does not, violating safety multiple times. 
}

\section{Conclusions}
In this article we presented a novel safety-critical control framework to handle unknown bounded disturbances for a broad class of nonlinear systems. We extended the method of backup CBFs to handle such disturbances by providing forward invariance conditions for a subset of a controlled invariant set governed by the disturbed system. 
We proved that enforcing these conditions guarantees safety for the disturbed system, and we demonstrated the effectiveness of the approach with two numerical simulation examples.

\bibliographystyle{ieeetr}
\bibliography{references}

\begin{thebibliography}{10}

\bibitem{ames_2017}
A.~D. Ames, X.~Xu, J.~W. Grizzle, and P.~Tabuada, ``Control barrier function based quadratic programs for safety critical systems,'' {\em IEEE Transactions on Automatic Control}, vol.~62, no.~8, pp.~3861--3876, 2017.

\bibitem{gurriet_online_2018}
T.~Gurriet, M.~Mote, A.~D. Ames, and E.~Feron, ``An {Online} {Approach} to {Active} {Set} {Invariance},'' in {\em 2018 {IEEE} {Conference} on {Decision} and {Control} ({CDC})}, pp.~3592--3599, IEEE, Dec. 2018.

\bibitem{gurriet_scalable_2020}
T.~Gurriet, M.~Mote, A.~Singletary, P.~Nilsson, E.~Feron, and A.~D. Ames, ``A scalable safety critical control framework for nonlinear systems,'' {\em IEEE Access}, vol.~8, pp.~187249--187275, 2020.

\bibitem{gurriet_thesis}
T.~Gurriet, {\em Applied Safety Critical Control}.
\newblock PhD thesis, California Institute of Technology, 2020.

\bibitem{jankovic_robust_2018}
M.~Jankovic, ``Robust control barrier functions for constrained stabilization of nonlinear systems,'' {\em Automatica}, vol.~96, pp.~359--367, Oct. 2018.

\bibitem{garg_robust_2021}
K.~Garg and D.~Panagou, ``Robust {Control} {Barrier} and {Control} {Lyapunov} {Functions} with {Fixed}-{Time} {Convergence} {Guarantees},'' in {\em 2021 {American} {Control} {Conference} ({ACC})}, pp.~2292--2297, IEEE, May 2021.

\bibitem{breeden_robust_2023}
J.~Breeden and D.~Panagou, ``Robust {Control} {Barrier} {Functions} under high relative degree and input constraints for satellite trajectories,'' {\em Automatica}, vol.~155, p.~111109, Sept. 2023.

\bibitem{shaw_cortez_control_2021}
W.~Shaw~Cortez, D.~Oetomo, C.~Manzie, and P.~Choong, ``Control {Barrier} {Functions} for {Mechanical} {Systems}: {Theory} and {Application} to {Robotic} {Grasping},'' {\em IEEE Transactions on Control Systems Technology}, vol.~29, pp.~530--545, Mar. 2021.

\bibitem{das2023robustcontrolbarrierfunctions}
E.~Daş, S.~X. Wei, and J.~W. Burdick, ``Robust control barrier functions with uncertainty estimation,'' {\em arXiv}, 2023.

\bibitem{issf_og}
M.~Z. Romdlony and B.~Jayawardhana, ``On the new notion of input-to-state safety,'' in {\em 2016 IEEE 55th Conference on Decision and Control (CDC)}, pp.~6403--6409, 2016.

\bibitem{Issf_ames}
S.~Kolathaya and A.~D. Ames, ``Input-to-state safety with control barrier functions,'' {\em IEEE Control Systems Letters}, vol.~3, no.~1, pp.~108--113, 2019.

\bibitem{issf_application_ames}
A.~Alan, A.~J. Taylor, C.~R. He, A.~D. Ames, and G.~Orosz, ``Control barrier functions and input-to-state safety with application to automated vehicles,'' {\em IEEE Transactions on Control Systems Technology}, vol.~31, no.~6, pp.~2744--2759, 2023.

\bibitem{TEZUKA_issf}
I.~Tezuka, T.~Kuramoto, and H.~Nakamura, ``Input-to-state constrained safety zeroing control barrier function and its application to time-varying obstacle avoidance for electric wheelchair,'' {\em IFAC-PapersOnLine}, vol.~55, no.~41, pp.~44--51, 2022.
\newblock 4th IFAC Workshop on Cyber-Physical and Human Systems CPHS 2022.

\bibitem{ames_adaptiveCBF}
A.~J. Taylor and A.~D. Ames, ``Adaptive safety with control barrier functions,'' in {\em 2020 American Control Conference (ACC)}, pp.~1399--1405, 2020.

\bibitem{xiao_adaptiveCBF}
W.~Xiao, C.~Belta, and C.~G. Cassandras, ``Adaptive control barrier functions,'' {\em IEEE Transactions on Automatic Control}, vol.~67, no.~5, pp.~2267--2281, 2022.

\bibitem{lopez_adaptive_robustCBF}
B.~T. Lopez, J.-J.~E. Slotine, and J.~P. How, ``Robust adaptive control barrier functions: An adaptive and data-driven approach to safety,'' {\em IEEE Control Systems Letters}, vol.~5, no.~3, pp.~1031--1036, 2021.

\bibitem{Taylor2019LearningFS}
A.~J. Taylor, A.~W. Singletary, Y.~Yue, and A.~D. Ames, ``Learning for safety-critical control with control barrier functions,'' in {\em Conference on Learning for Dynamics \& Control}, 2019.

\bibitem{lindemann2024learningrobustoutputcontrol}
L.~Lindemann, A.~Robey, L.~Jiang, S.~Das, S.~Tu, and N.~Matni, ``Learning robust output control barrier functions from safe expert demonstrations,'' {\em IEEE Open Journal of Control Systems}, vol.~3, pp.~158--172, 2024.

\bibitem{taylor_robust_dataDriven}
A.~J. Taylor, V.~D. Dorobantu, S.~Dean, B.~Recht, Y.~Yue, and A.~D. Ames, ``Towards robust data-driven control synthesis for nonlinear systems with actuation uncertainty,'' in {\em 2021 60th IEEE Conference on Decision and Control (CDC)}, pp.~6469--6476, 2021.

\bibitem{fernando_data-driven}
F.~Castañeda, J.~J. Choi, B.~Zhang, C.~J. Tomlin, and K.~Sreenath, ``Pointwise feasibility of {Gaussian} process-based safety-critical control under model uncertainty,'' in {\em 2021 60th IEEE Conference on Decision and Control (CDC)}, pp.~6762--6769, 2021.

\bibitem{emam_data_driven}
Y.~Emam, P.~Glotfelter, S.~Wilson, G.~Notomista, and M.~Egerstedt, ``Data-driven robust barrier functions for safe, long-term operation,'' {\em IEEE Transactions on Robotics}, vol.~38, no.~3, pp.~1671--1685, 2022.

\bibitem{abate_robust_monotonicity}
M.~Abate and S.~Coogan, ``Computing robustly forward invariant sets for mixed-monotone systems,'' in {\em 2020 59th IEEE Conference on Decision and Control (CDC)}, pp.~4553--4559, 2020.

\bibitem{abate_enforcing_2020}
M.~Abate and S.~Coogan, ``Enforcing {Safety} at {Runtime} for {Systems} with {Disturbances},'' in {\em 2020 59th {IEEE} {Conference} on {Decision} and {Control} ({CDC})}, pp.~2038--2043, IEEE, Dec. 2020.

\bibitem{cosner_measurement-robust_2021}
R.~K. Cosner, A.~W. Singletary, A.~J. Taylor, T.~G. Molnar, K.~L. Bouman, and A.~D. Ames, ``Measurement-robust control barrier functions: Certainty in safety with uncertainty in state,'' in {\em 2021 {IEEE}/{RSJ} International Conference on Intelligent Robots and Systems ({IROS})}, pp.~6286--6291, {IEEE}, 2021.

\bibitem{chen2021backupcontrolbarrierfunctions}
Y.~Chen, M.~Jankovic, M.~Santillo, and A.~D. Ames, ``Backup control barrier functions: Formulation and comparative study,'' 2021.

\bibitem{khalil2002nonlinear}
H.~Khalil, {\em Nonlinear Systems}.
\newblock Pearson Education, Prentice Hall, 2~ed., 2002.

\bibitem{freeman_robust_1996}
R.~A. Freeman and P.~Kokotović, {\em Robust Nonlinear Control Design}.
\newblock Birkhäuser Boston, 1996.

\bibitem{contraction_bullo}
F.~Bullo, {\em Contraction Theory for Dynamical Systems}.
\newblock Kindle Direct Publishing, {1.1}~ed., 2023.

\end{thebibliography}

\end{document}